\documentstyle[aps,prl,epsfig,floats,twocolumn]{revtex}

\def\@cite#1{{\footnotesize $^{#1}$}}

\begin{document}


\twocolumn[\hsize\textwidth\columnwidth\hsize\csname @twocolumnfalse\endcsname

\title{\Large Thermoelectric Response Near the Density Driven Mott Transition}

\author{Gunnar P\'alsson and Gabriel Kotliar }
\address{Serin Physics Laboratory, 
Rutgers University, PO Box 849, Piscataway NJ, 08855.}

\date{May 25, 1998}

\maketitle

\begin{abstract}
We investigate the thermoelectric response of correlated electron
systems near the density driven Mott transition using the dynamical
mean field theory.
\end{abstract}

\pacs{PACS Numbers: 75.20.Hr, 72.15.Jv, 71.55.Jv}

] 


Thermoelectric effects which are responsible for the direct conversion
of heat energy into electrical energy and vice versa have recently
received renewed attention \cite{Mahan:1997}.  So far optimal
thermoelectric response has been achieved in traditional doped
semiconductors but recent improvements in both the synthesis of
correlated electron systems and the theoretical methods for treating
them have generated new interest in the thermoelectric properties of
correlated materials.

The thermoelectric performance of a material, reflects its ability to
convert applied voltages into temperature gradients while minimizing
the irreversible effects of Joule heating and thermal conduction.
This can be quantified by the so called dimensionless figure of merit
denoted here by $Z_{T}T$:
\begin{equation}
\label{eq:figureofmerit}
{Z_{T}} T ={{S^2 \sigma T} \over \kappa+\kappa_{L} }.
\end{equation}
Here $\sigma$ is the electrical conductivity, $S$ the thermopower (or
Seebeck coefficient) and $T$ the temperature, $\kappa$ is the
electronic contribution to the thermal conductivity and $\kappa_{L}$
is the lattice thermal conductivity.  In weakly correlated systems the
Seebeck coefficient can be interpreted as the logarithmic derivative
of the conductivity with respect to the Fermi energy.  In strongly
correlated electron systems S is a more difficult quantity to
interpret, and is this the subject of this letter.

We consider a system of fermions doped away from a Mott insulating
state, where the magnetic correlations are weak so that the magnetism
is not the driving force behind the metal to insulator transition.
This situation is realized experimentally in titanate and vanadate
perovskite compounds \cite{Tokura:1993a}.  For example,
La$_{1-x}$Y$_{x}$TiO$_3$ is ferromagnetic for $x$ near 1 and
anti-ferromagnetic at small values of $x$ but in all these compounds the
N\'eel and Curie temperatures are quite small, of the order of 130 K
and less \cite{Okimoto:1995} so we focus on the paramagnetic phase of
the doped Mott insulator.

The experimental motivation for our study was mainly provided by the
work of Y. Tokura's group which has demonstrated that in the class of
ternary perovskites, one can control to a large extent various
parameters such as orbital degeneracy, bandwidth, and carrier
concentration to provide an experimental realization of the filling
driven Mott transition \cite{Tokura:1993a}.

To approach this problem we use the Dynamical Mean Field Theory (DMFT)
\cite{Metzner:1989,Georges:1992a,Jarrell:1992,Pruschke:1995,Georges:1996},
which has successfully described many aspects of the physics of
three dimensional transition metal oxides.  The goal of our study is
to understand qualitatively the thermoelectric response of a doped
Mott insulator and to derive explicit formulas for the transport
coefficients which are valid at very low and high temperatures. Using
these insights and numerical results we discuss the figure of merit of
this particular kind of materials.
An early  numerical calculation of the Seebeck coefficient
in the large d Hubbard model appeared
in
ref. \cite{Pruschke:1995}.

We consider perfect periodic solids described by the $N$-fold
degenerate Hubbard model:
\begin{equation}
\label{eq:Hubbard-Hamiltonian}
H = -\sum_{\langle ij\rangle\sigma}t_{ij}c_{i\sigma}^{+}c_{j\sigma} +
{U \over 2}\sum_{j\sigma\neq\sigma'} n_{j\sigma}n_{j\sigma'}
-\mu\sum_{j\sigma}n_{j\sigma}
\end{equation}
and ignore electron-phonon interactions.  The index $\sigma$ can be
thought of as a spin or an orbital index, and will run from 1 to
$N$. The single band case corresponds to $N=2$.  We now summarize the
relevant aspects of the DMFT.  The transport properties are obtained
from a Green's function with a frequency dependent but momentum
independent self energy:
\begin{equation}
G(k,\omega) = G(\epsilon_{k},\omega) = 
{1 \over \omega + \mu - \epsilon_{k} - \Sigma(\omega)}
\end{equation}
where $\epsilon_{k}$ is the dispersion relation.  The self-energy, is
computed by solving an Anderson impurity model in a bath described by
a hybridization function $\Delta (i \omega) $. Regarded as a
functional of the hybridization function it obeys the self consistency
condition \cite{Georges:1992a,Jarrell:1992}.
\begin{equation}
{1 \over i\omega+\mu-\Delta(i\omega)-\Sigma(i\omega)} =
\int d\varepsilon G(\varepsilon,i\omega)D(\varepsilon)
\end{equation}
where $D(\varepsilon)$ is the bare density of states, $D(\varepsilon)
= \sum_{k}\delta(\varepsilon-\epsilon_{k}).  $ The transport
coefficients that govern the electrical and thermal responses of the
model are given in terms of current-current correlation functions.
Within the DMFT they reduce to averages over the spectral density
$\rho (\epsilon, \omega)$ \cite{Pruschke:1995}:
\begin{equation}
\label{eq:trancofs}
\sigma = {e^{2} \over  T}A_{0},
\quad
S = {-k_{B} \over e} {A_{1} \over A_{0}},
\quad
\kappa= k_{B}^{2}(A_{2} - {A_{1}^{2} \over A_{0}}).
\end{equation}
where
\begin{eqnarray}
\label{eq:corrfunc}
A_{n} &=& {\pi \over \hbar V}\sum_{k,\sigma} \int d \omega
\rho_{\sigma}(k,\omega)^2({\partial \epsilon_k \over \partial k_x})^2
(-T{\partial f(\omega) \over \partial \omega})(\beta\omega)^{n}
\nonumber \\ &=& {N\pi \over \hbar k_{B}}
\int_{-\infty}^{\infty}d\omega
d\varepsilon{\rho^{2}(\varepsilon,\omega) (\omega\beta)^{n}\over
4\cosh^{2}({\beta\omega \over 2})}\Phi(\varepsilon).
\end{eqnarray}
In this expression the relevant information about the bare band
structure is contained in the spectral density, $\rho$, and the
transport function $\Phi$,
\begin{equation}
\Phi(\varepsilon) = {1 \over V}\sum_{k} ({\partial \epsilon_{k} \over
\partial k_x})^2 \delta(\varepsilon-\epsilon_{k}).
\end{equation}

The DMFT  expressions bear superficial  similarity
with recent  work of Mahan and Sofo
\cite{Mahan:1996}, if we identify their transport 
function 
which we 
denote 
$\Sigma_{\mbox{\tiny{MS}}}(\varepsilon)$
as:
\begin{equation}
\Sigma_{\mbox{\tiny{MS}}}(\omega+\mu) = \left({N\pi \over \hbar}\right)\int
d\varepsilon\rho^{2}(\varepsilon,\omega)\Phi(\varepsilon).
\end{equation}
Ref. \cite{Mahan:1996} stressed the relevance of correlated materials
to thermo-electricity and suggested that the optimal response is
obtained when $\Sigma_{\mbox{\tiny{MS}}}$ takes the form of a delta
function like peak at the Fermi level.  The DMFT allows us to derive
explicit expressions for $\Sigma_{\mbox{\tiny{MS}}}$ starting from
microscopic Hamiltonians.  Near the Mott transition, the low energy
spectral function does assume a delta like form at low temperatures.
We find however, that the optimum of the figure of merit is achieved
at high temperatures, when the quasi-delta-function-like resonance at
the Fermi level is absent.

In electronic systems near a Mott transition there are two widely
separated energy scales: the bare bandwidth D, which sets the scale
for incoherent excitations and the coherence temperature
$T_{\mbox{\small{coh}}}$ where Fermi-liquid-like properties begin to
be observed.  $T_{\mbox{\small{coh}}}$ vanishes as we approach half
filling.  As a result when we decrease the temperature, starting from
very high temperatures we expect two crossovers to take place, one
when $k_{B}T \approx D$ and the other when $T \approx
T_{\mbox{\small{coh}}}$.  We assume in this analysis that the
interaction energy U is much larger than the bare bandwidth, i.e.  we
are close to the the $U=\infty$ limit.

At very low temperature the Fermi liquid picture is applicable and
momentum space offers a natural description of the transport
processes.  The only states that contribute come from a narrow region
around the Fermi surface, which in the local mean-field picture shows
up as a narrow Kondo like resonance near the Fermi level.  At very
high temperatures the entire Brillouin zone is equally important and a
real-space picture of the transport is more appropriate.  The
transport is then entirely due to incoherent motion of charge
carriers.  We discuss these two asymptotic regimes below.

{\bf Low temperature regime:} In this regime we use Fermi liquid ideas
to parameterize the transport coefficients in terms of a few
parameters, which contain all the effects of the interactions:
$\gamma = Im\Sigma(0)$ is the scattering rate at the Fermi level, $Z =
(1-{\partial Re\Sigma \over \partial \omega}|_{\omega = 0})^{-1}$ is
the quasiparticle residue at the Fermi-level and $\tilde{\mu} =
\mu-Re\Sigma(0)$ is the effective chemical potential.  Close to the
Mott transition there is only one effective scale controlling the
low-temperature physics and therefore the scattering rate behaves as
$A(k_{B}T)^{2}/{Z^2D}$ where A is a dimensionless constant
\cite{Goetz:1995}.  Performing a low-temperature expansion of equation
(\ref{eq:corrfunc}) and substituting into (\ref{eq:trancofs}), we
obtain:
\begin{equation}
\sigma = {Ne^{2}DZ^{2}\Phi(\tilde{\mu})E_{0} \over 2A(k_{B}T)^{2}}
\qquad
\kappa =  {k_{B}NDZ^{2} \over 2A(k_{B}T)}\Phi(\tilde{\mu})E_{2}
\end{equation}
We find that the thermopower and the figure of merit do not depend on
$\gamma$ and are given by
\begin{eqnarray}
\label{eq:fltep}
S &=& -{k_{B} \over e}\left({k_{B}T \over Z} {d \ln\Phi(\tilde{\mu})
\over d\tilde{\mu}}\right){E_{2} \over E_{0}} \\
\label{eq:flfom}
Z_{T}T &=& {E_{2} \over E_{0}}\left( {k_{B}T \over
Z}{d\ln\Phi(\tilde{\mu}) \over d\tilde{\mu}} \right)^{2}.
\end{eqnarray}
Here the numbers $E_{n}$ are given by,
\begin{equation}
E_{n} = \int_{-\infty}^{\infty}{x^{n}dx \over 4\cosh^{2} \left({x
\over 2}\right) \left[1+\left({x \over \pi}\right)^{2}\right]}.
\end{equation}
Numerical calculation gives $E_{0} = 0.82$ and $E_{2} = 1.75$.
Note that the Wiedeman-Franz law holds here as it does for transport
dominated by impurity scattering.  However, since the scattering rate
is energy dependent around the Fermi-surface the ratio of $\sigma T$
and $\kappa$ is not equal to the classical Lorenz number.  The
physical content of equation (\ref{eq:flfom}) is transparent.
Correlations enhance the figure of merit relative to that of a
non-interacting system with the same density of states by a factor of
$1 \over Z^2$ which in this context can be thought of as the square of
the mass enhancement. This factor is expected to be large and in fact
diverges as we approach the density driven Mott transition in
La$_{1-x}$Sr$_{x}$TiO$_{3}$ \cite{Tokura:1993a}.  Note however that
expression (\ref{eq:flfom}) is only valid in the low-temperature
regime because it is restricted to $\beta ZD \gg 1$.  Thus the figure
of merit will be very low in the low-temperature regime unless the
logarithmic derivative of the transport functions becomes appreciable.
Unlike the density of states however, the transport function does not
have any Van Hove singularities and the only singular points in the
logarithmic derivative are at the band edges where the transport
function vanishes.

{\bf High temperature regime:} 
To describe   this region we  observe  that the spectral
function shifted by the value of the chemical potential
\begin{equation}
\widetilde{\rho}(\epsilon,\omega) = \rho(\epsilon,\omega-\mu)
\label{eq:rhotilde}
\end{equation}
converges to a well-defined, shape centered around $\omega = 0$ as the
temperature tends to infinity.  This agrees with a rigid-band
interpretation of the lower Hubbard band, except that the carriers in
this band are completely incoherent near the band edge.  The
high-temperature behavior of the chemical potential can also be found
analytically and is given by: $\beta\mu = \ln({n \over N(1-n)})$.  We
can now obtain the leading high-temperature behavior of the transport
coefficients by inserting the scaling form (\ref{eq:rhotilde}) with
$\widetilde{\rho}$ temperature independent and expanding to lowest
order in $\beta$ the equations for the coefficients $A_{n}$.  The
results are parameterized in terms of the moments
\begin{equation}
\gamma_{m} = {a \over D^{m+1}}\int d\epsilon d\omega \omega^{m}
\widetilde{\rho}^{2}(\epsilon,\omega)\Phi(\epsilon)
\end{equation}
which we evaluate numerically.  To leading order in $\beta$ we find:
\begin{eqnarray}
\sigma &=& \left({e^{2} \over a\hbar}\right)\pi N(D\beta)\gamma_{0}
{{n \over N}(1-n) \over ({n \over N}+(1-n))^{2}}
\\
\label{eq:heikes}
S &=& \left({k_{B} \over e}\right)\ln({n \over N(1-n)}) \\
\label{eq:varmi}
\kappa &=& \left({k_{B}D \over a\hbar}\right)\pi
N(D\beta)^{2}\gamma_{2} {{n \over N}(1-n) \over ({n \over
N}+(1-n))^{2}}.
\end{eqnarray}
Near the Mott transition the density and degeneracy dependence of the
moments $\gamma_{0}$ and $\gamma_{2}$ is given by: $\gamma_{m} =
\widetilde{\gamma}_{m}(1-n+{n \over N})^{2}$ where the
$\widetilde{\gamma}_{m}$'s are constants, $\widetilde{\gamma}_{0}
\approx 0.05$ and $\widetilde{\gamma}_{2} \approx 0.01$.  The factor
in parenthesis is simply the total integrated spectral weight of the
Green's function.  The high-temperature equation that we get for the
thermopower corresponds to the well known Heikes formula
\cite{Chaikin:1976} generalized for degeneracy $N$.  Comparison with
numerical solutions of the dynamical mean field equations, reveals
that for the resistivity, the high temperature expansion formula is
quite accurate over a very wide temperature range and breaks down only
at temperatures of the order of $T_{\mbox{\small{coh}}}$. This is
surprising, since the high temperature expansion is a priori only
valid for $\beta D \ll 1$.  This result gives some insight into
the origin of the linear
resistivity which was   observed  in early studies of 
the Hubbard model in infinite dimensions
\cite{Jarrell:1994}.  Furthermore if one keeps the next term in the
high-temperature expansion of the thermopower, one can fit the
numerical data over a comparable region.  Notice that the thermopower
close to the Mott transition is hole-like which agrees with the
picture of holes in a paramagnetic spin background \cite{Oguri:1990}.

The high-temperature expression for the figure of merit is to lowest
order in $\beta$ given by:
\begin{equation}
Z_{T}T = {\pi\widetilde{\gamma_{0}}
\ln^{2}\left({n \over N(1-n)}\right) n(1-n)
\over {\kappa_{L} \over \kappa_{D}}+
\pi\widetilde{\gamma_{2}}(D\beta)^{2}n(1-n)},
\end{equation}
where $\kappa_{D} = {k_{B}D \over a\hbar}$.  Here the lattice
contribution to the thermal conductivity has been included since the
electronic contribution tends to zero in the high temperature region.
As temperature increases the figure of merit increases monotonically
to a constant value that goes linearly with the bandwidth, $D$, of the
system.  At any finite temperature however the equation above gives a
figure of merit that increases monotonically with $D$ to a maximum
which is at a bandwidth larger than $k_{B}T$ and thus outside the
region of validity of our formula.  Thus we conclude that the optimum
figure of merit is obtained when the bandwidth is of the order of the
temperature.

\begin{figure}[ht]
\epsfig{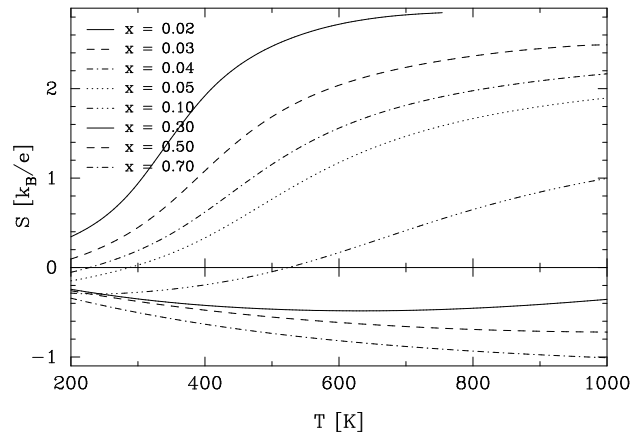}
\vspace{6pt}
\caption{Thermopower vs. temperature for several values of doping in
the La$_{1-x}$Sr$_{x}$TiO$_{3}$ system.}
\label{fig:thermopower}
\end{figure}

We now turn to quantitative calculations of the thermopower in the
intermediate temperature range which is the range most pertinent to
experiments and possible applications.  The best characterized system,
in the class of materials that we seek to describe is the
La$_{1-x}$Sr$_{x}$TiO$_3$ system with $x$ small.  We model this system
with a Hubbard model on a three dimensional hypercubic lattice with
half bandwidth $D = 0.5$ eV and interaction strength $U = 2.0$ eV in
the Mott insulating end of the series ($x=0$).  We take into account
the $x$ dependence of the bandwidth by using the fact that the Ti-O-Ti
bond-angle, $\theta$, changes with doping.  The bandwidth then depends
on $\theta$ through $D(\theta) = D(180^{\circ})\cos^{2}\theta$.  To
select the bond-angles corresponding to a given doping we use data for
ab-plane bond-angles from \cite{Sunstrom:1992}.  The results from the
calculations of S, using IPT, with this choice of parameters is
displayed in Fig. \ref{fig:thermopower}.
We notice here that in the high temperature end of the data the values
of the thermopower are actually close to what Heikes formula predicts
thus indicating that the high temperature thermopower can be used as a
rough estimate of the carrier density.  At lower temperature we see a
marked decrease in the magnitude of the thermopower which heralds the
transition into the Fermi-liquid regime where the thermopower is small
and electron-like.  We note however that for doping values lower than
0.1 the thermopower is already hole-like above 300K which indicates
that the quasiparticle peak has almost disappeared and that the
carriers are completely incoherent. This regime does not appear for
smaller values of the interaction U when the system is below the Mott
transition.  The thermopower near the Mott transition therefore seems
to be a sensitive indicator of the character of the carriers
(i.e. localized vs. itinerant).

\begin{figure}[ht]
\epsfig{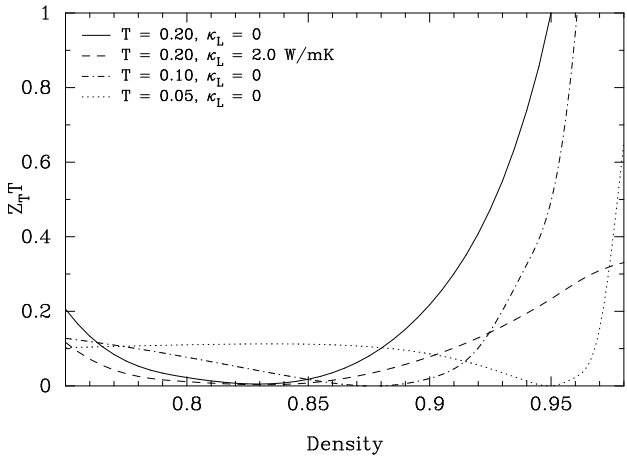}
\vspace{6pt}
\caption{Thermopower vs. temperature for several values of
doping. Calculated with IPT using U = 4.0.}
\label{fig:fom}
\end{figure}

Since bandwidth and carrier concentration, can be controlled to a
large degree in this class of materials we use our results to assess
the prospects for a large figure of merit in this class of systems.
We are not aware of any measurements of the thermal conductivity in
the titanate compounds so for the purpose of obtaining an order of
magnitude estimate of the figure of merit we use a value of $2.0$W/mK
for $\kappa_{L}$ based on measurements carried out recently in the
manganite oxides \cite{Visser:1997},

The results from the numerical calculation of the figure of merit in
the intermediate temperature region are displayed in Fig.\
\ref{fig:fom}.  The data for T = 0.05 and T = 0.10 were obtained with
iterated perturbation theory (IPT) modified for finite doping
\cite{Kajueter:1996b}.  For the T = 0.20 data we use the infinite $U$
non crossing approximation (NCA).  This approach has also been shown
to give results in good agreement with exact methods at high
temperatures \cite{Pruschke:1993b,Costi:1996}.
The behavior of the figure of merit can be understood qualitatively as
follows.  It vanishes at a density $n_{0}(T)$ where the Seebeck
coefficient vanishes. At very high-temperature this occurs at $n_{0} =
{N \over 1+N}$ according to equation (\ref{eq:heikes}).  As
temperature is lowered $n_{0}(T)$ moves towards half filling, and is
given by the condition $T_{\mbox{\small{coh}}}(n) \approx T$.  This is
due to the formation of the coherent quasi-particles which give a
negative contribution to the thermopower.  For densities lower than
$n_{0}(T)$ the system is Fermi-liquid-like and we find a low figure of
merit, closer to the Mott transition at temperature higher than
$T_{\mbox{\small{coh}}}(n)$ the figure of merit is larger.

The lowest temperature results show little density dependence below
$n_{0}$.  This is because both the logarithmic derivative of the
transport function and the quasiparticle residue depend essentially
linearly on the doping and therefore the figure of merit varies
little with density.  The T = 0.1 data seems to be
displaying a similar trend as the T = 0.05 data but the system is
essentially out of the Fermi-liquid regime at that temperature.

To conclude, we have investigated the thermoelectric coefficient and
the thermoelectric figure of merit near the density driven Mott
transition.  We provided simple expressions for the transport
coefficients of this model, obtaining a qualitative understanding of
the thermoelectric coefficient.  In the light of our results, a large
figure of merit in this class of systems seems rather unlikely.  It
would require very small doping, and a very small value of the bare
half-bandwidth parameter D, to be able to access the high temperature
regime $T > D$ where the figure of merit is of order unity.  We have
argued however, that the thermopower is a sensitive probe of the
degree of itineracy of the carriers, and an experimental investigation
of the thermoelectric response in the region where large mass
enhancement has been observed \cite{Tokura:1993a} is highly desirable.

\noindent {\bf  Acknowledgements}:

We acknowledge useful discussions with Ekkehard Lange, Henrik Kajueter, 
A. Ramirez and P.B. Littlewood.  This work was supported by the
NSF under grant DMR 95-29138.

\appendix

\end{document}